\documentclass[twocolumn,aps]{revtex4}
\usepackage{epsfig}
\usepackage{psfig}
\usepackage{amsmath}

\newcommand{\be}{\begin{equation}}
\newcommand{\ee}{\end{equation}}
\newcommand{\bea}{\begin{eqnarray}}
\newcommand{\eea}{\end{eqnarray}}

\begin{document}
\title{Template fitting and the large-angle CMB anomalies}
\author{Kate Land$^1$ and Jo\~{a}o Magueijo$^{2,1}$ }
\address{$^1$ Theoretical Physics Group, Imperial College, Prince Consort Road,
London SW7 2BZ, UK\\
$^2$Perimeter Institute for Theoretical Physics, 31 Caroline St. N.,
Waterloo, N2L 2Y5, Canada}

\begin{abstract}
We investigate two possible explanations for the large-angle
anomalies in the Cosmic Microwave Background (CMB): an
intrinsically anisotropic model and an inhomogeneous model.
We take as an example of the former a Bianchi model (which
leaves a spiral pattern in the sky) and of the latter a background
model that already contains a non-linear long-wavelength plane wave
(leaving a stripy pattern in the sky).
We make use of an adaptation of the
``template'' formalism, previously designed to detect galactic
foregrounds, to recognize these patterns and produce confidence
levels for their detection. The ``corrected'' maps, from which
these patterns have been removed, are free of anomalies, in
particular their quadrupole and octupole are not planar and their
intensities not low. We stress that although the ``template''
detections are not found to be statistically significant they do correct
statistically significant anomalies.
\end{abstract}

\pacs{PACS Numbers: *** }
\keywords{
cosmic microwave background - Gaussianity tests - statistical isotropy}
\date{\today}

\maketitle

\section{Introduction}\label{intro}

Since the release of the first year data from NASA's Wilkinson
Microwave Anisotropy Probe \citep{wmap} there have been extensive
studies of the Cosmic Microwave Background (CMB) maps. Generally
the results impressively support the favoured $\Lambda$CDM
cosmological model, though a lack of power on the largest scales
has produced much debate (see {\it e.g.}, \cite{efstat, bridle,
biel, contaldi, teg, riaz, slosar2}). Of further interest are
studies that have detected evidence for significant
departures from the fundamental cosmological assumption that the
universe is isotropic and homogeneous
\citep{hbg,erik1,erik3,teg,vielva,copilet,evil,dodec,ral,us,tesse,hansen}.
Thus it appears the large angle CMB multipoles are anomalous in two
seemingly distinct ways. Firstly their power $C_\ell$ is
abnormally low. Secondly they have an improbable directionality
revealed by the fact that for a certain $z$-axis' orientation one
$m$-mode absorbs most of the power, for $\ell=2,...,5$
\citep{evil}. Naively one may expect the two features to be
related: the absence of power in all possible $m$ modes may be why
the observed average $C_\ell$ is so low. But this has to be made
rigorous. Such is the purpose of this letter: could it be that
accounting for the large-angle directional multipoles accounts for
the observed low $C_\ell$?

$\Lambda$CDM theory matches the power spectrum very well on most
scales \citep{wmap}, but the strong non-Gaussianity and/or
anisotropy of the low multipoles suggests a competition between
two processes: Gaussian isotropic random fluctuations and a
deterministic process. One possibility for the latter is intrinsic
cosmic anisotropy (as encoded in the Bianchi models) which imparts
a spiral pattern in the CMB as studied
in~\cite{barrow,pedro,kogut,tesse}.
Another is a background
model that already contains a non-linear long-wavelength plane wave.
This would leave a striking stripy pattern in the sky (a ``Jupiter'',
if the polar axis is defined suitably). The matter may then be addressed by
regarding these patterns as a ``contamination'' template, and
attempting to ``correct'' the data for its coupling to the
template, thereby purifying the underlying Gaussian process. The
formalism has been developed to deal with galactic foregrounds but
may be adapted to this situation.

We therefore seek a template capable of explaining the preferred
axis in the CMB, and ask the question: could the corrected map
reveal higher $C_\ell$s, thus linking the two issues? Note that
this is highly non-trivial. Usually when one corrects a map for a
``foreground'' the power in the corrected map is {\it smaller}.
However it could also be that the deterministic process acted
destructively. For this to be possible the template would have to
have a pattern complementary to the observed data, {\it i.e.},
contain power in (at least some of the) $m$s which fail to reveal
it in the data; in addition this power would have to have acted
destructively upon the underlying Gaussian process, so that only
one $m$ mode survived in the data.

The presence of a template is more likely to add power than to
remove it, so to explain the low-$\ell$ low power this way one
relies in part on a chance alignment of phases between the
Gaussian fluctuations and the template. Still, in this paper we
explore this approach with two different models. In both cases,
once we correct the data, and are left with a map with higher
low-multipole power $C_\ell$, and no preferred direction as seen in
\cite{evil}.

In Section~\ref{form} we outline our method for selecting favoured
templates. Then in~\ref{temp} we introduce
the templates of our two specific models, and in
Section~\ref{results} we explain how we selected the data and
record our results. We discuss the
implications of these in Sections~\ref{sims} and~\ref{discuss}.

\section{The Formalism}\label{form}

Our work is modelled on that of~\cite{tesse} but with an important
extension. We start by reviewing~\cite{tesse}. Suppose the sky
${\mathbf d}$ may be modelled as the sum of a Gaussian process
${\mathbf g}$ and a non-Gaussian ``template'', ${\mathbf t}$. Then
the Maximum Likelihood Estimation (MLE) problem may be set up by
introducing a coupling parameter $\alpha$ so that ${\mathbf
d}={\mathbf g}+\alpha {\mathbf t}$. The usual MLE problem now
concerns ${\mathbf d}-\alpha {\mathbf t}$, that is: \bea
-2\ln {\cal L}&=&\chi^2+\ln \det {\mathbf M}\\
\chi^2&=&({\mathbf d}-\alpha {\mathbf t})^{\rm T}{\mathbf M}^{-1}
({\mathbf d}-\alpha {\mathbf t})\label{quad} \eea (we have ignored
an additive constant in the first expression). Here ${\mathbf M}$
is the covariance matrix of ${\mathbf g}$ (which may include
noise) and the data vectors may be either temperatures in pixel
space, or a set of spherical harmonic coefficients $a_{\ell m}$.
In the latter case the data is complex, satisfying ``reality''
constraints; the concomitant modifications to the formalism are
trivial. For full sky maps  the $a_{\ell m}$ are the preferred
formulation.

In general we have a joint estimation problem for  $\alpha$ and
the unknowns parameterizing ${\mathbf M}$ (which can be binned
$C_\ell$ or the half-dozen parameters encoding ``cosmic
concordance''). However it is often the case that the range of
$\ell$s which decides the ${\mathbf M}$ problem and that for the
template are disjoint. For example the geometry of the Universe
$\Omega$, or the amount of baryonic matter $\Omega_b$ are mainly
decided by the Doppler peaks, whereas we are interested in
templates producing large-angle non-Gaussian features (say $\ell=2 -5$;
see~\cite{evil}). One may then decouple the two problems: find
${\mathbf M}$ ignoring the template and then solve for $\alpha$
using this solution for ${\mathbf M}$.

\subsection{Parameter Estimation}

The MLE problem for $\alpha$ then becomes the problem of
minimizing $\chi^2$. The solution is \be\label{almin}
\alpha_{min}={{\mathbf t}^{\rm T}{\mathbf M}^{-1}{\mathbf d}\over
{\mathbf t}^{\rm T}{\mathbf M}^{-1}{\mathbf t}} \ee By
construction $\alpha$ is a Gaussian variable. Its variance is
\be\label{dal} \delta \alpha^2={1\over {\mathbf t}^{\rm T}{\mathbf
M}^{-1}{\mathbf t}} \ee Two comments are in order. Firstly notice
that this formalism wants the template to be like the data. If a
mysterious theory was found predicting ${\mathbf t}={\mathbf d}$,
then $\alpha=1$ and the MLE problem tells us that the Gaussian
process is in fact absent \footnote{This is a bit oversimplistic,
since the error bar $\delta \alpha$ may in fact be very large. We
thank Andrew Jaffe for pointing this out to us.}. This is not in
line with common sense: we do know that there {\it is} a Gaussian
process. Secondly this procedure always produces a corrected map
${\mathbf d}-\alpha {\mathbf t}$ with lower chi-squared than the
original map ${\mathbf d}$. This is fine if one suspects
contamination because of an abnormally high chi-squared, but what
if the observed chi-squared is too low? In that case the corrected
map is even more abnormal.

 As is usually the
case, these quibbles boil down to the choice of parameters and
priors used. We have failed to impose a suitable condition {\it
enforcing} the presence of a Gaussian process, so that the
corrected maps have a reduced chi-squared close to 1. If we are in
$a_{\ell m}$ space, for example, we are assuming a uniform prior
in $a_{\ell m}$. The peak of the distribution is always  at
$a_{\ell m}=0$, so if ${\mathbf t}$ is such that this condition
can be satisfied that is precisely what the formalism does (a zero
chi-squared is the highest probability configuration if one
assumes uniform priors in $a_{\ell m}$). Seen in another way, if
we have a deterministic template competing with a Gaussian process
on equal footing, the formalism will always try to suppress the
random Gaussian process in favour of the certainties of the
template.

To address this issue we use an alternative function of the data
in calculating the likelihood. We base the likelihood on the
probability distribution of the $\ln(\chi^2)$ (this rather than
$\chi^2$ produces the correct power spectrum estimator). By
changing variables the likelihood is given by \be -2\ln {\cal
L}=\chi^2+\ln \det {\mathbf M}+D\ln(\chi^2)\label{like}\ee where
$D$ is the number of degrees of freedom (i.e. the number of
pixels, or the number of $a_{\ell m}$ being considered). Its
maximization with respect to $\alpha$ now leads to: \be {\partial
\chi^2\over
\partial \alpha}{\left( {D\over \chi^2} -1\right)}=0 \ee i.e.
either the first factor is zero (equivalent to the problem of
minimizing the $\chi^2$ solved above), or $\chi^2=D$, which means
the reduced chi-squared is 1. This is simple to solve, and
involves finding the roots to the quadratic equation (\ref{quad}).
The solution is \be\label{alphapm} \alpha_\pm={\chi^2_{dt}\pm
{\sqrt{\chi^4_{dt}-\chi^2_{t}( \chi^2_{d}-D)}}\over \chi^2_{t}}
\ee where \bea
\chi^2_{t}&=&{\mathbf t}^{\rm T}{\mathbf M}^{-1}{\mathbf t}\\
\chi^2_{d}&=&{\mathbf d}^{\rm T}{\mathbf M}^{-1}{\mathbf d}\\
\chi^2_{dt}&=&{\mathbf t}^{\rm T}{\mathbf M}^{-1}{\mathbf d} \eea
This formalism for correcting the map generalizes that
of~\cite{tesse}, while reducing to it in the appropriate
circumstances. If the quadratic does not have real solutions for
$\alpha$ (because the minimum of $\chi^2(\alpha)$ is above $D$)
then the problem reduces to~\cite{tesse}. This is the case where
the chi-squared of the data is too high, so that correcting it
entails reducing it. In this case the two formalisms agree.
However if the data's chi-squared is too low then our formalism
takes over. The quadratic then has real solutions, and the MLE
problem is solved by solutions $\alpha_\pm$ given by
(\ref{alphapm}).

We note that this approach could be conceptually improved by
factorizing the likelihood $\ell$ by $\ell$ and imposing uniform
priors on an appropriate function of {\it each} $\chi^2_\ell$. Minimising the
likelihood with respect to $\alpha$ would now lead to the problem of solving
\be\label{fullprob}
\sum_\ell {\partial \chi^2_\ell\over
\partial \alpha}{\left( {D\over \chi^2_\ell} -1\right)}=0 \ee
However, this is in practice impossible to solve, especially for
large $\ell_{\rm max}$. Therefore, we are solving for the simpler case
of $\chi^2=\sum_\ell\chi^2_\ell$.

\subsection{Model Comparison}

We have found the necessary correction to make for a given
template. Perhaps more important is to find the preferred template
- {\it i.e.}, the preferred model. The Bayesian approach to model
comparison is to marginalise over the parameters of a model, to
find the probability of data given the model (whatever the
parameters), as this relates to the probability of the model given
the data. We have been working with $P(\alpha,t)$, and above we
discuss the specific $\alpha$ solutions that maximise this for a
given template. We now wish to compare the likelihood of different
templates, that is we require just $P(t)$ - a marginalization over
$\alpha$. The standard approach to this (see for e.g \cite{AJ,
slosar}) is to absorb the uncertainties of $\alpha$ into the
covariance matrix. That is \be \chi^2={\mathbf d}^{\mathbf T}
{\mathbf M'}^{-1} {\mathbf d} \ee for \be {\mathbf
M'}^{-1}=\lim_{\sigma\to\infty}({\mathbf M} + \sigma^2{\mathbf
t}^{\mathbf T}{\mathbf t})^{-1}\ee where ${\mathbf M}$ is the
usual covariance matrix associated with the $a_{\ell m}$s as used
above. Using the Sherman-Morrison-Woodbury \citep{AJ} formula we
find \be {\mathbf M'}^{-1}={\mathbf M}^{-1} - {\mathbf
M}^{-1}{\mathbf t}({\mathbf t}^{\mathbf T}{\mathbf M}^{-1}{\mathbf
t}){\mathbf t}^{\mathbf T}{\mathbf M}^{-1}\ee and therefore \be
\chi^2=\chi^2_{d} -
\frac{(\chi^2_{dt})^2}{\chi^2_{t}}\label{newchi}\equiv\chi^2_{d}
-\Gamma^2\ee
Assuming uniform priors on our templates, our model comparison now
involves maximising the likelihood (\ref{like}) using the
new expression for $\chi^2$ and the matrix ${\mathbf M'}$.
In~\cite{slosar} there is a discussion of  how this method of marginalising over
$\alpha$ inside the covariance matrix is equivalent to ignoring
the data that correlates with the template when working out the
$\chi^2$.

The likelihood function (\ref{like}) has its maximum at
$\chi^2=D$. The effective chi-squared from a template is given by
Eq.~(\ref{newchi}) and we see it is always less than $\chi^2_d$.
Therefore, if the initial chi-squared of the data is low: $\chi^2
< \chi^2_d < D$, the preferred template is that which maximises
the $\chi^2$ (minimises $\Gamma$) so as to bring it as close to
$\chi^2_d$ as possible, the optimal solution being $\Gamma = 0$.
Conversely, if the data has a high chi-squared our formalism
reduces to that of \cite{tesse}, where we maximise $\Gamma$ so as
to reduce the $\chi^2$ to as close to D as possible.

\section{The Templates}\label{temp}
We consider two prototypes of violation of isotropy and homogeneity:
Bianchi models and plane-wave inhomogeneous models.

\subsection{Bianchi models}
Bianchi models are well-known homogeneous, anisotropic generalizations
of Friedmann-Robertson-Walker models. They may be used as a competitor
to standard cosmology, constraining violations of isotropy while still
assuming homogeneity. In these models the CMB is never isotropic, even before
fluctuations are added on. Of particular interest are Bianchi $VII_h$ models,
for which an asymmetric spiral pattern is imprinted upon the sky.
These models were extensively examined in~\cite{tesse}
with reference to asymmetries in the power spectrum and anomalous hot-spots.
We have reproduced these results with our programs.
One of the deficiencies of the fits found in~\cite{tesse} (recognized by
the authors) is that they require $\Omega\neq 1$. This would almost
certainly lead to a bad fit at high $\ell$ even if the large-angle corrected
map is improved. For this reason here we shall restrict ourselves to
$\Omega=1$ models.

We follow the parametrisation of the Bianchi $VII_h$ model from
\cite{barrow}. These parameters are $\Omega$,
$x=\sqrt{h/(1-\Omega)}$, handedness, and the Euler angles
$(\phi,\theta,\psi)$\footnote{We use the "x-convention", where the
Euler angles $(\phi,\theta,\psi)$ represent first a rotation by
$\phi$ around the z-axis, then by $\theta$ around the x-axis, then
by $\psi$ around the z-axis (again).}. We refer the reader to
\cite{barrow} for a fuller explanation. We fit templates with
$\Omega=1$, and explore the range $x\in[0.1,10]$, with both left
and right handedness, over the total range of Euler angles.

\subsection{Plane-wave cosmology}
Inhomogeneous cosmologies are either very simple
or very complicated~\cite{inhom}. We consider a model in which
a plane wave in the gravitational potential $\Phi$ is part and parcel of the background model
(i.e. the unperturbed, zeroth order cosmological model, before the standard Gaussian isotropic
scale-invariant adiabatic fluctuations are added to it).

An outstanding wave might be the result of several processes. It could
be for example the hallmark of non-trivial topology~\cite{gomero, dodec, riaz}.
Non-trivial topologies
may be ruled out on the grounds that they would imprint repeat patterns
in the CMB sky, e.g. matching circles in antipodal locations for the most basic
``slab topology''. This constraint is only applicable if the fundamental domain
is smaller than the last scattering surface.

Larger-than-the-horizon domains have two effects upon the density plane waves riding the
homogeneous model. Firstly modes must fit into the domain, leading
to a discretization. But they also introduce a scale in the
problem, the size of the domain, $L$. Hence the usual naive arguments for
scale-invariance no longer apply and $k^3\Phi^2$ may actually be a function
of $kL$ without introducing an arbitrary scale in the problem.
We speculate that even if $L$ is sufficiently large that one may
neglect mode discretization inside the horizon, the fundamental
mode $k=2\pi/L$ will be very intense, and dominate over whatever
underlying (scale-invariant) Gaussian process. Interestingly in~\cite{gomero}
it was shown that multipole alignments, like those observed \citep{copilet,teg,evil},
are produced in certain non-trivial topologies.

Another context in which strong long-wavelength modes have been
studied is the work of \cite{kolb}. Here (controversial) claims
were made that super horizon scale fluctuations may be causing
the acceleration of the universe. It would be interesting to investigate
further the exact length/intensity ratio required for such a model
to work (should it work at all). The impact upon the CMB would then fall
under the present study.

Regardless of these two possible motivations it is interesting to
examine the evidence for one such wave in the CMB, and it is
this more phenomenological approach that will be followed in
our paper.
We will search for evidence of a dominating ${\mathbf k}$-mode
with wavelength $\lambda \in [0.5,10]$ in units of the
diameter of the last scattering surface.
Our parameters for this model are the wavelength
$\lambda$, the phase $\rho$, and the Euler angles
$(\phi,\theta,\psi)$ (we actually do not need to concern ourselves
with the first rotation $\phi$ about the z-axis as the wave is
cylindrically symmetric).

\section{Results}\label{results}

We use full sky maps, specifically we look at the
cleaned map of \cite{teg0} as this map has been shown to have
interesting features~\cite{evil}. In our covariance matrix we
include only the theoretical power spectrum terms from WMAP
\cite{wmap}, and ignore noise (we are considering only the
low-$\ell$s). We include a 24$^o$ beam, and consider the
$\ell$-range: 2 - 5 (32 degrees of freedom). For this $\ell$ range
this map finds a $\chi^2_d=27.06 < 32$, and so the formalism
outlined in Section~\ref{form}
requires that we select the template that minimizes $\Gamma$.

We use this $\ell$-range for 2 reasons. Firstly, two significant
and different anomalies have been seen on these scales; a
preferred direction and low power. Secondly, this is the relevant $\ell$-range
for the templates we are examining.
Considering an $\ell$ range not covered by a template is pointless
when the purpose is to assess the effect a template will have on
the map.


\begin{table}
\begin{tabular}{|l|ll|ll|ll|}
                \hline
           $\ell$&Before&&Bianchi&&Wave&\\
      & ${\mathbf n}$ & $m$ & corrected && corrected &\\
                \hline
         2&(-103,59)&2&(156,8)&0&(-34,19)&0\\
         3&(-121,62)&3&(-26,14)&1&(-17,7)&1\\
        4&(-163,58)&2&(-156,59)&2&(85,11)&3\\
         5&(-93,49)&3&(-93,49)&3&(-93,48)&3\\
                \hline
        Mean inter angle  & 22.4  && 58.6 && 59.4 & \\
                \hline
\end{tabular}
\caption{The Axis of Evil behaviour before and after template
corrections. The ${\mathbf n}$ (in galactic coordinates $(l,b)$)
and $m$ are listed for the low multipoles $\ell=2-5$. The average
inter-angle of the 4 axis is shown. The uncorrected map finds a
strong alignment of axis, shown by the low average inter-angle.
(see Section~\ref{results}, and Equation~\ref{evileq}.)}
\label{eviltab}
\end{table}

\begin{figure}
\psfig{file=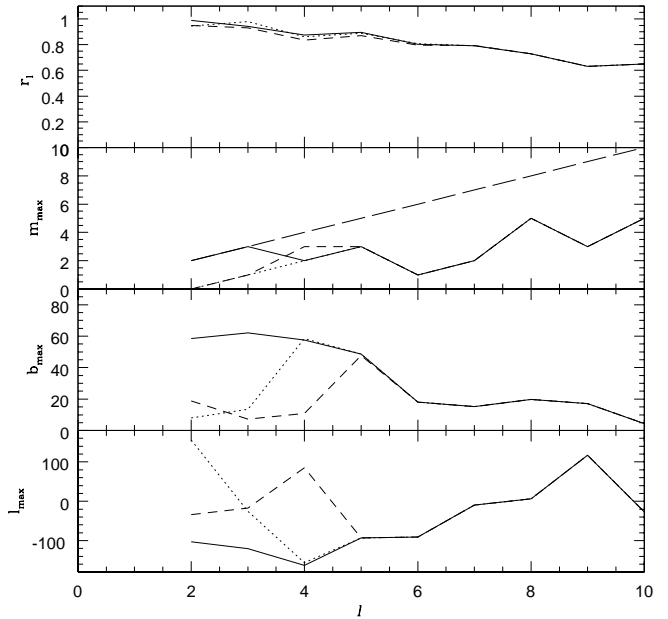,width=9.5cm}\caption{The Axis of Evil
behaviour, before (solid line) and after template corrections
(Bianchi corrected (dotted), Wave corrected (dashed)). See
Equation~\ref{evileq}. The top panel shows the ratio of power in
the preferred $m$, the second panel shows the $m$, and the bottom
2 panels display the axis ${\mathbf n}$ in galactic coordinates
$(l,b)$.}\label{new}
\end{figure}

\begin{figure}
\centering \hfill
\begin{minipage}{75mm}
\begin{center}
\psfig{file=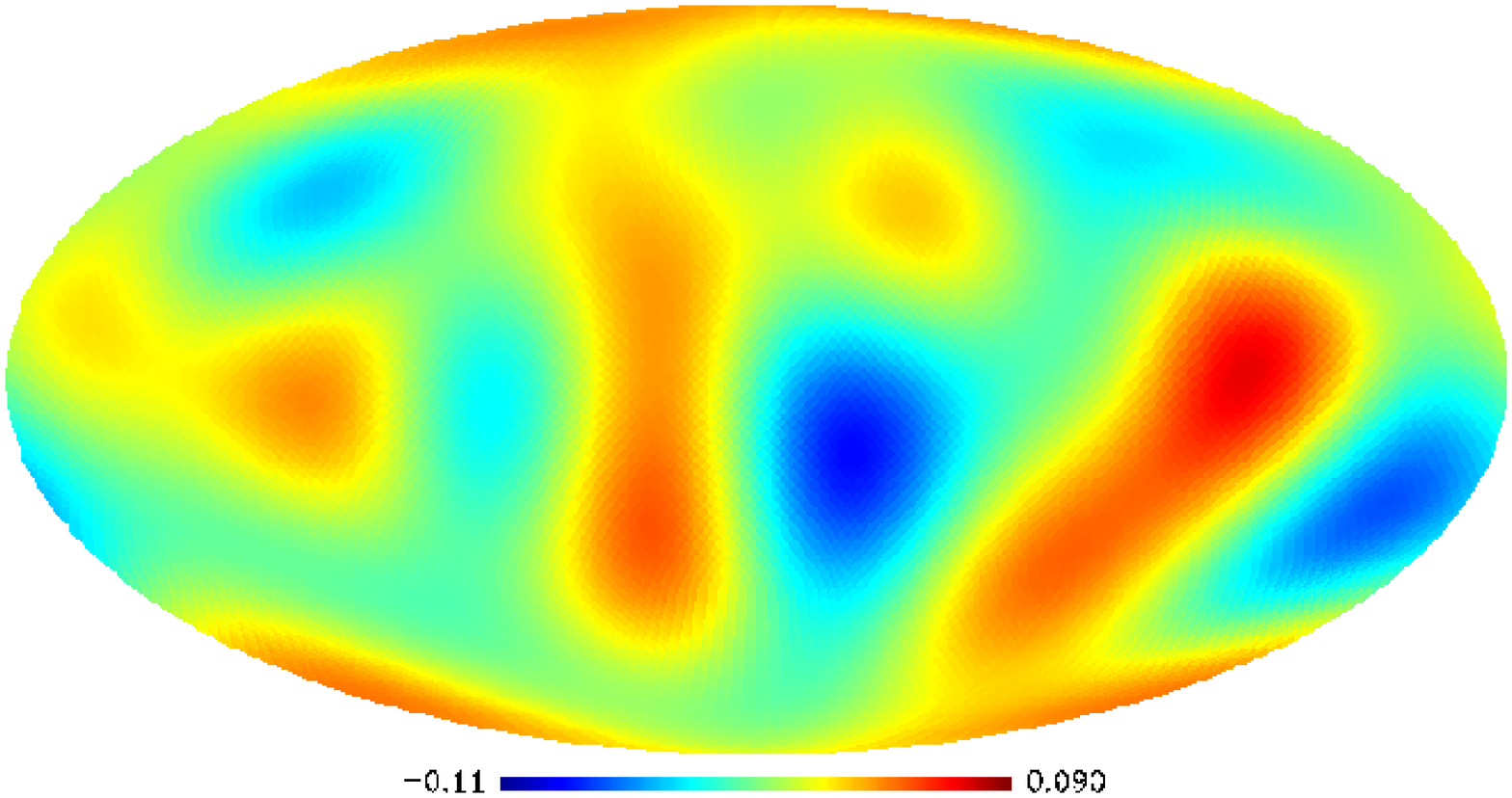,angle=90,width=7.5cm}\label{BB}\hfill
\psfig{file=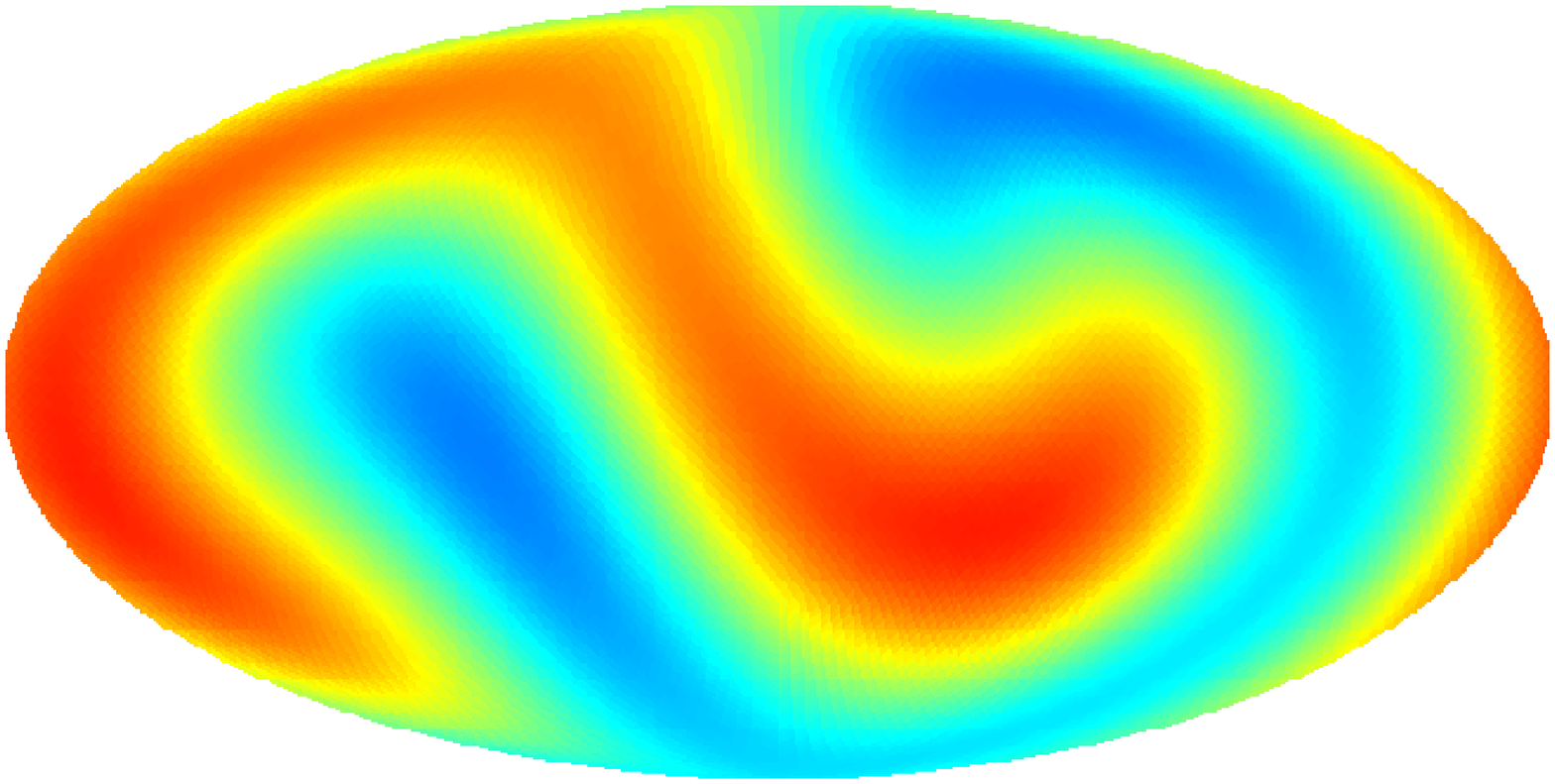,angle=90,width=7.5cm}\label{TB}\hfill
\psfig{file=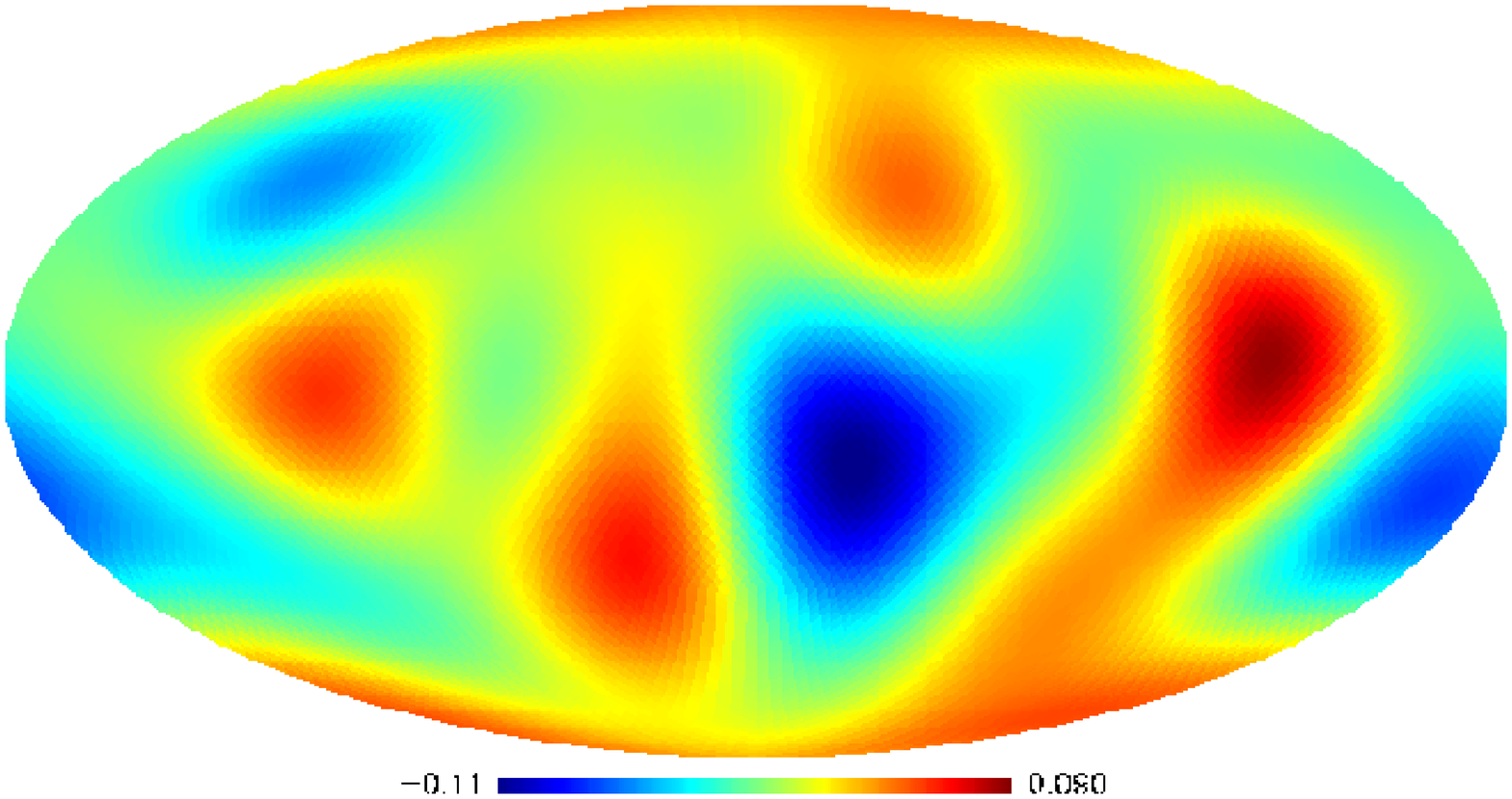,angle=90,width=7.5cm}\label{CB}\hfill
\end{center}
\end{minipage}
\hfill \caption{The map (mK) before (top) and after (bottom) it
has been corrected for the preferred Bianchi template (middle).
The template's power has been doubled for illustration purposes.}
\label{figB}
\end{figure}

\begin{figure}
\centering \hfill
\begin{minipage}{75mm}
\begin{center}
\psfig{file=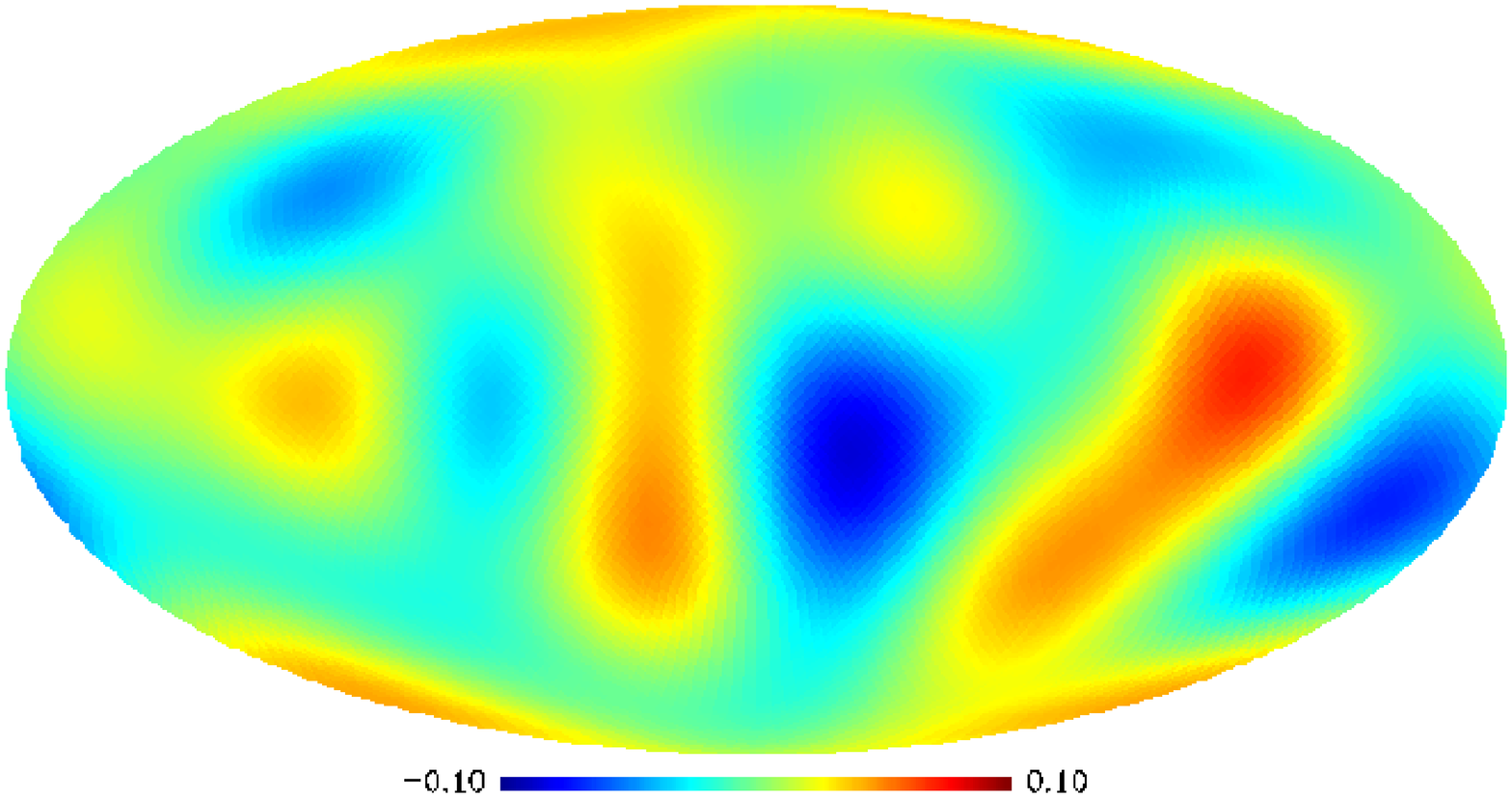,angle=90,width=7.5cm}\label{BW}\hfill
\psfig{file=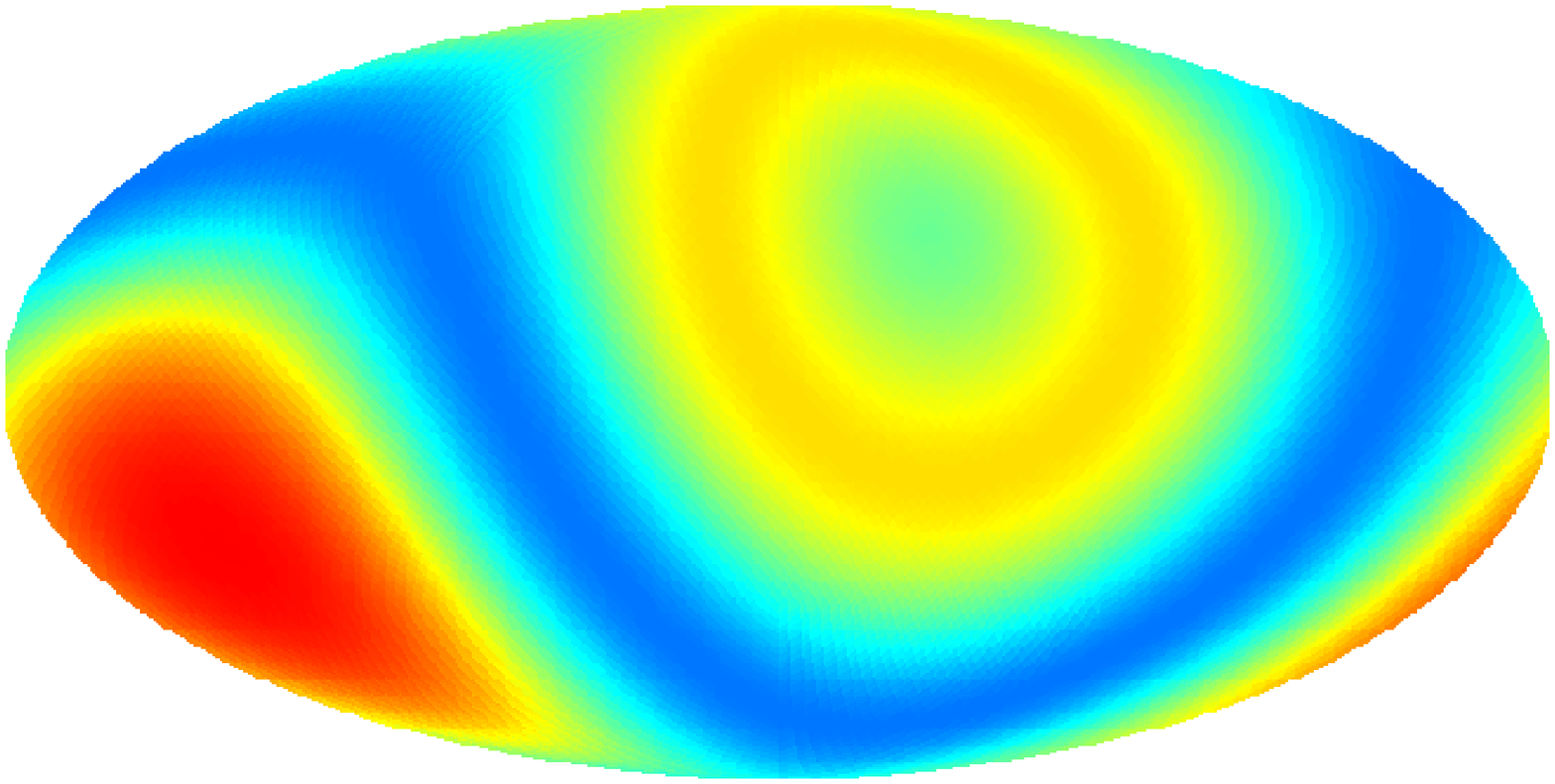,angle=90,width=7.5cm}\label{TW}\hfill
\psfig{file=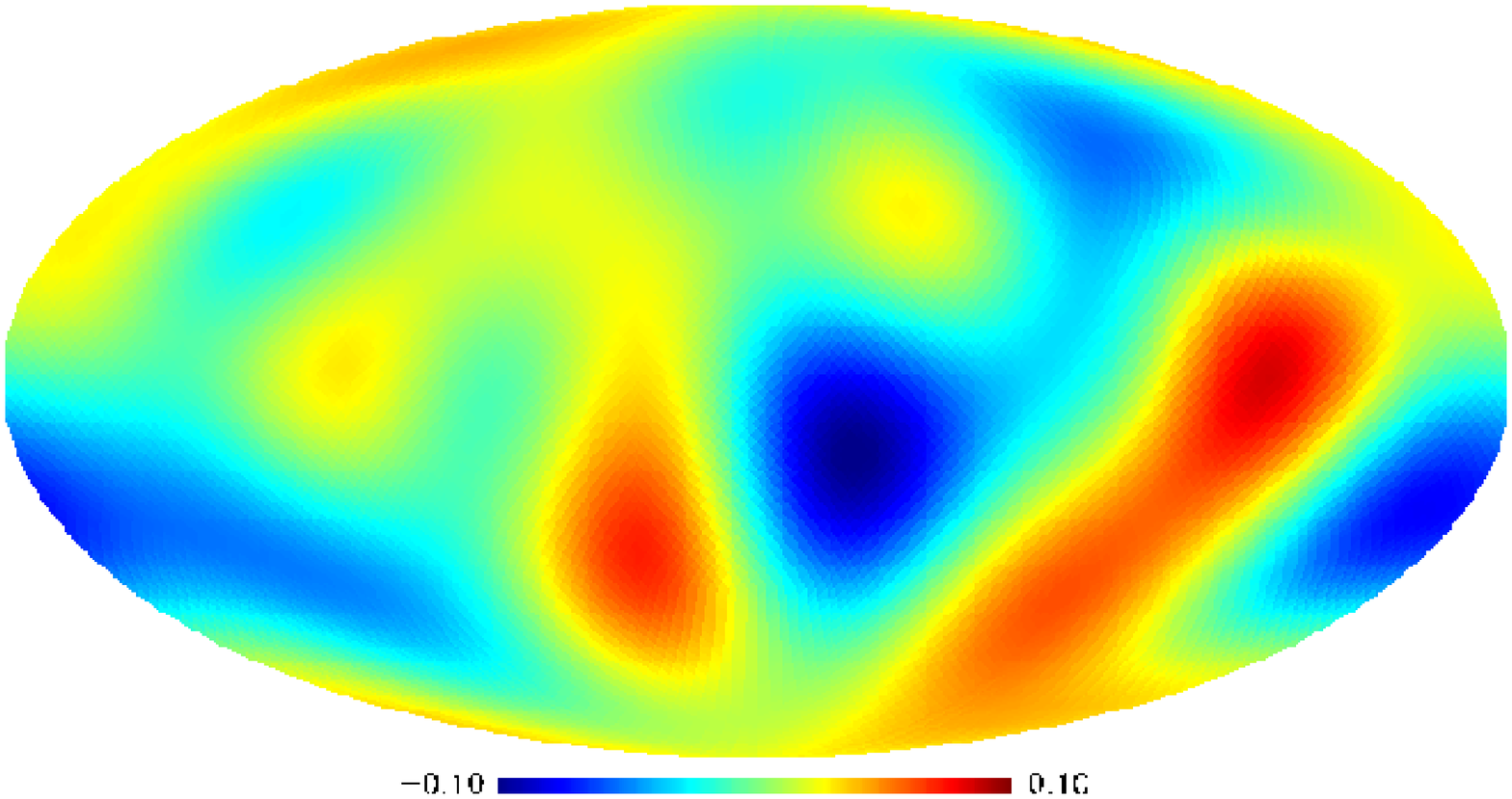,angle=90,width=7.5cm}\label{CW}\hfill
\end{center}
\end{minipage}
\hfill \caption{The map (mK) before (top) and after (bottom) it
has been corrected for the preferred Wave template (middle). The
template's power has been doubled for illustration purposes.}
\label{figW}
\end{figure}

We then minimize $\Gamma$ over the range of parameters for both
template types. It turns out that in neither case do we find one
clear optimal solution: there are numerous templates that find
$\Gamma\approx 0$ within the limitations of computational accuracy
and the rotation grid. This is due to the small $\ell$-range but
also due to the fact that the observed data is such a bad fit,
with a very low $\chi^2$. Note that we are limited to this narrow
$\ell$-range by our choice of models as this is where their power
lies.

Given the degeneracy of our solution we have to impose an
additional condition to select the preferred templates.
Considering that the full (unsolvable) problem requires
solving (\ref{fullprob}) we may break the degeneracy by
evaluating the $\chi^2_\ell$ and defining \be
\Delta=\sum_{\ell}\left(\frac{\chi^2_{\ell}}{2\ell+1} - 1\right)^2
\ee By minimising $\Delta$  we find the solution of the abridged
problem that's closest to the solution to the full problem.
Then the  corrected map will have  the
power spectrum closest to the expected power spectrum line. This
also breaks the degeneracy between the two $\alpha$ solutions
associated with each template (see equation~\ref{alphapm}).

We find that our preferred Bianchi  template has $x=1.4$ with
left handed vorticity, and Euler angles
$(\phi,\theta,\psi)=(-82,72,-62)$, and therefore an axis in the
direction $(l,b)=(-62,18)$ in galactic coordinates.
We find the preferred large Wave template has $\lambda = 0.96$ and
$\rho = 44^o$, in the direction $(l,b)=(-38,30)$ (equivalently a
wave in the opposite direction with $(l,b)=(142,-30)$ and phase
$\pi-\rho$). That is, the preferred wave is just slightly smaller
than the size of the CMB.

We examined the ``Axis of Evil'' behavior of the corrected maps
\citep{evil}, and results are in Table~\ref{eviltab} and
Figure~\ref{new}. We define a direction for each multipole by
finding the $m$ and the direction ${\bf n}$ that maximises \be
r_\ell= \frac{C_{\ell m}(\bf n)} {(2\ell +1){\hat C}_\ell}
\label{evileq}\ee where $C_{\ell 0}=|a_{\ell 0}|^2$, $C_{\ell
m}=2|a_{\ell m}|^2$ for $m>0$ (notice that 2 modes contribute for
$m\neq 0$) and $(2\ell +1){\hat C}_{\ell }=\sum_m |a_{\ell m}|^2$.
This finds the axis for which a multipole is most dominated by
just one $m$-mode. As can be seen in Table~\ref{eviltab}, the
uncorrected map finds similar directions for the $\ell=2,3,4,5$
multipoles (as reported in \cite{evil}) - a significant departure
from isotropy. We see that the corrected maps find no such
anisotropic behaviour. The average inter-$\ell$ angles found are
perfectly consistent with isotropy as compared to simulations.

\begin{figure*}
\centering \hfill
\begin{minipage}{55mm}
\begin{center}
\psfig{file=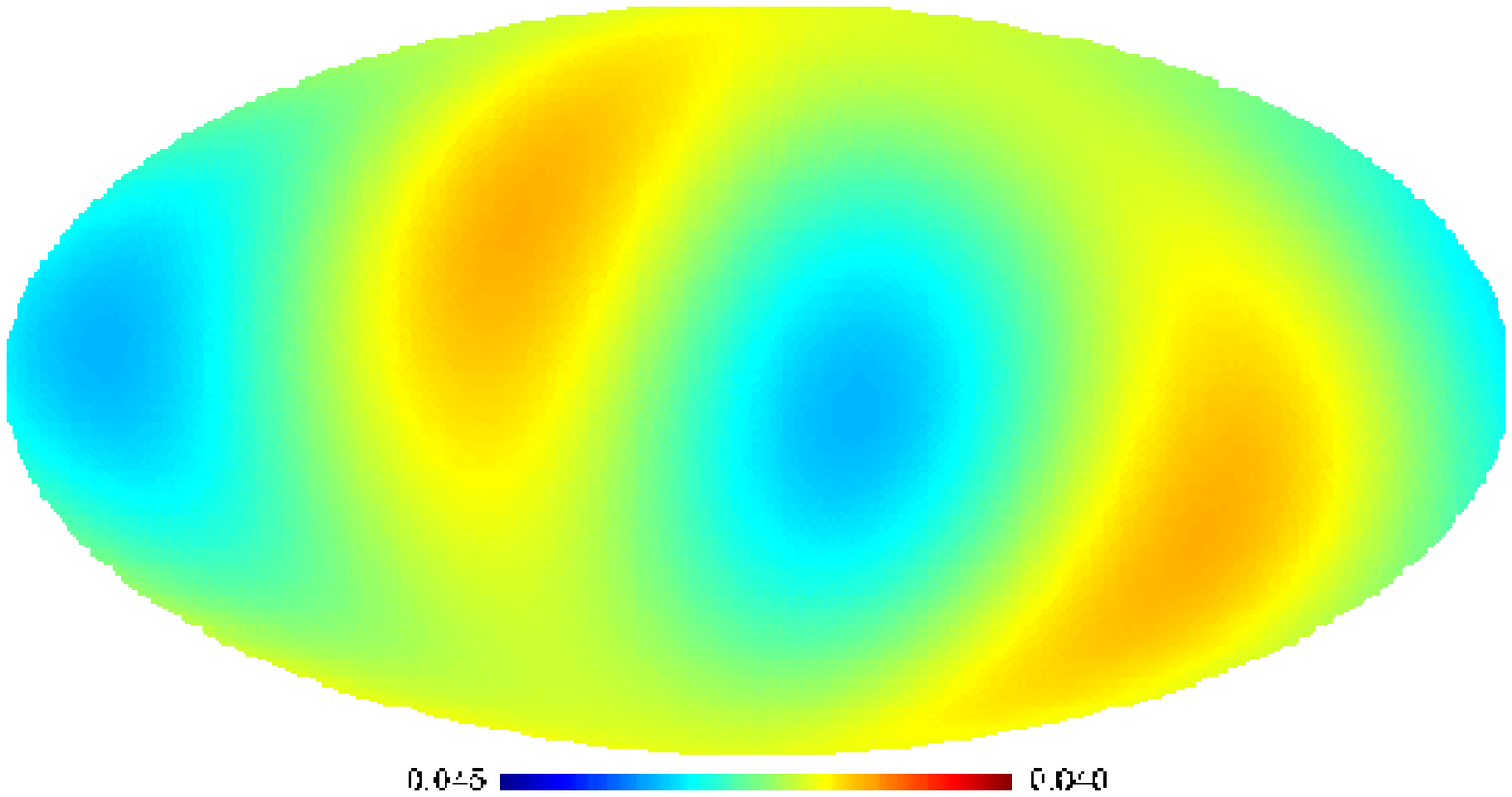,angle=90,width=5.5cm}\label{in2}\hfill
\psfig{file=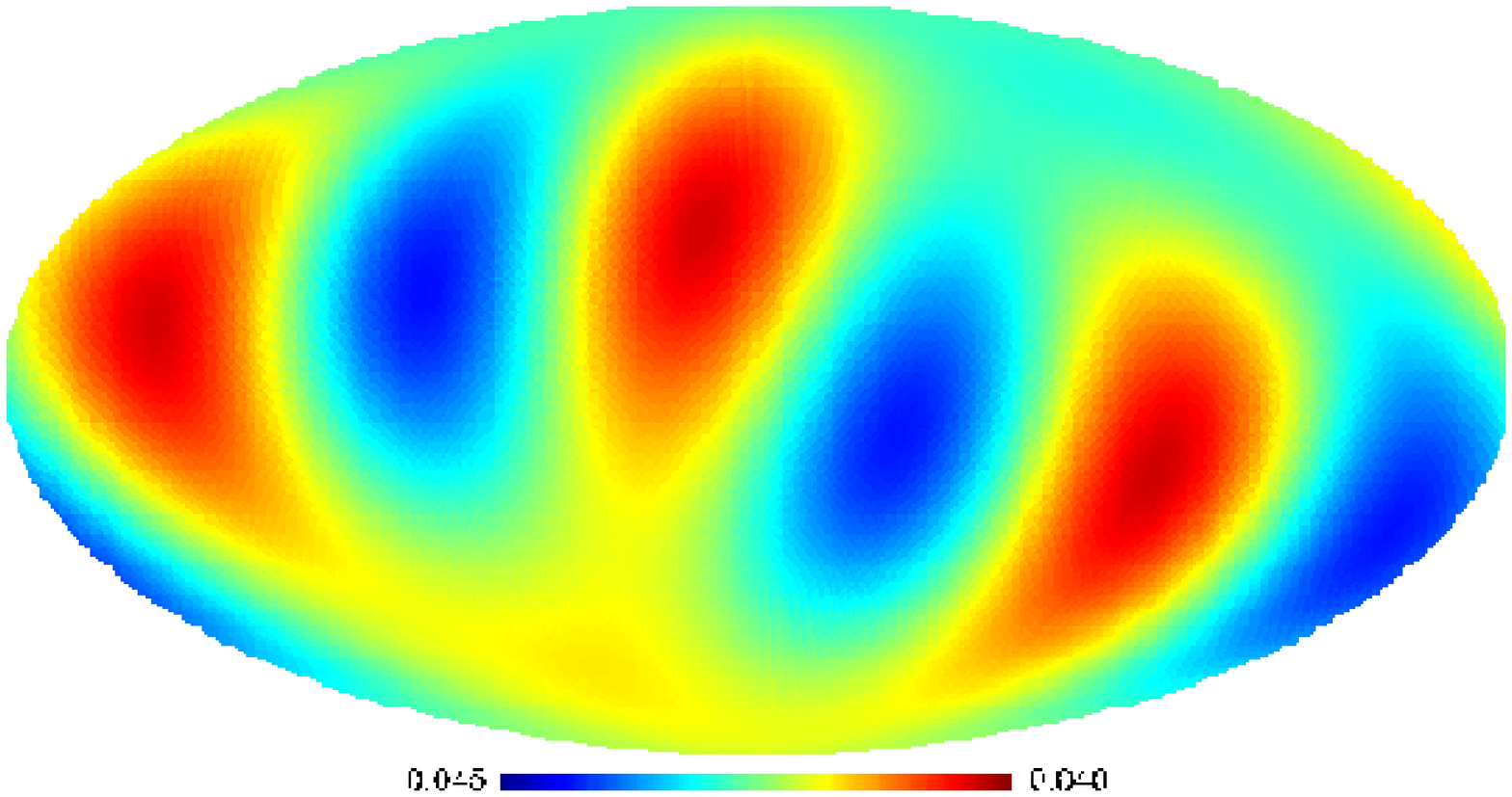,angle=90,width=5.5cm}\label{in3}\hfill
\end{center}
\end{minipage}%
\hfill
\begin{minipage}{55mm}
\begin{center}
\psfig{file=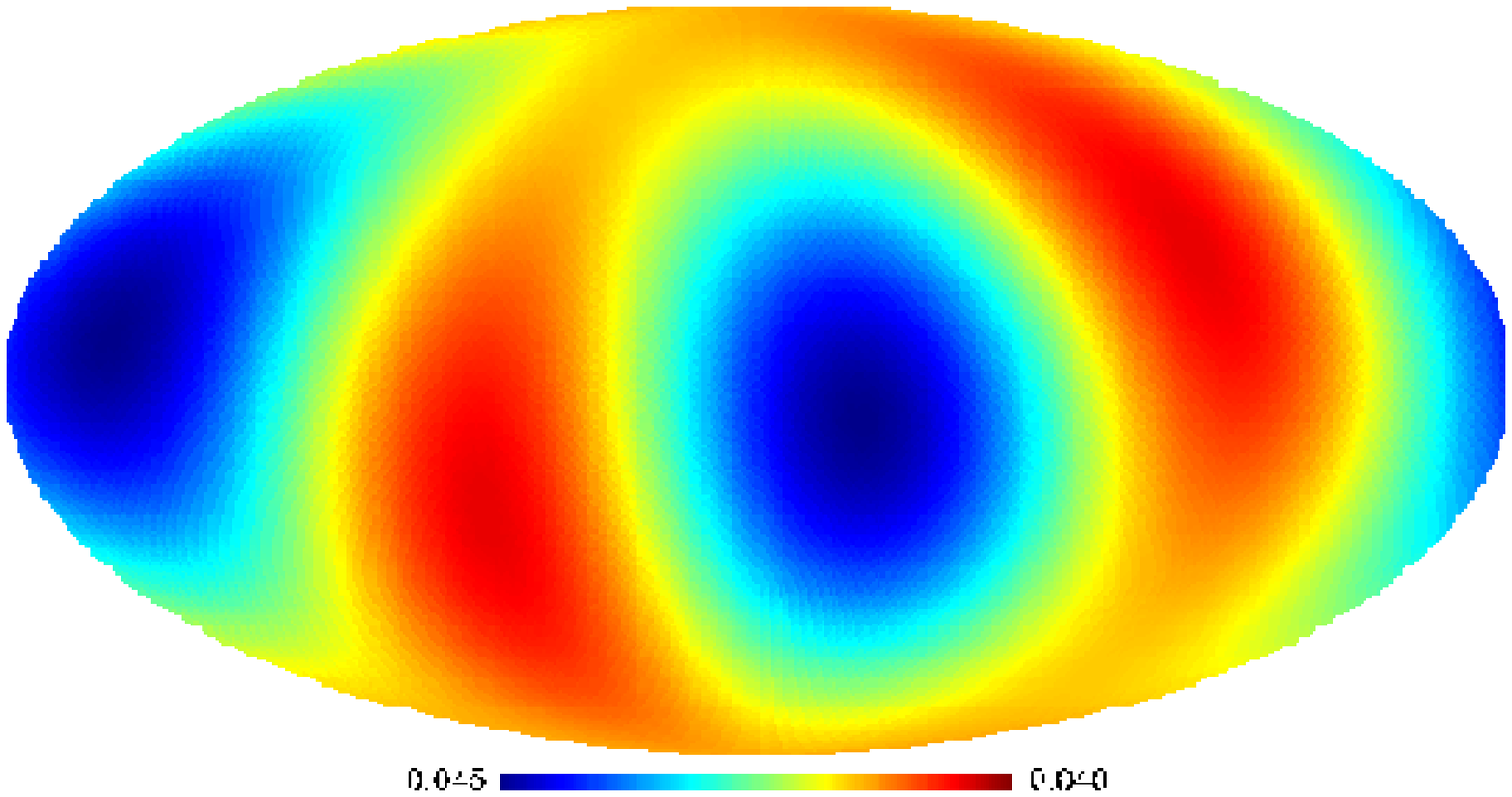,angle=90,width=5.5cm}\label{out2B}\hfill
\psfig{file=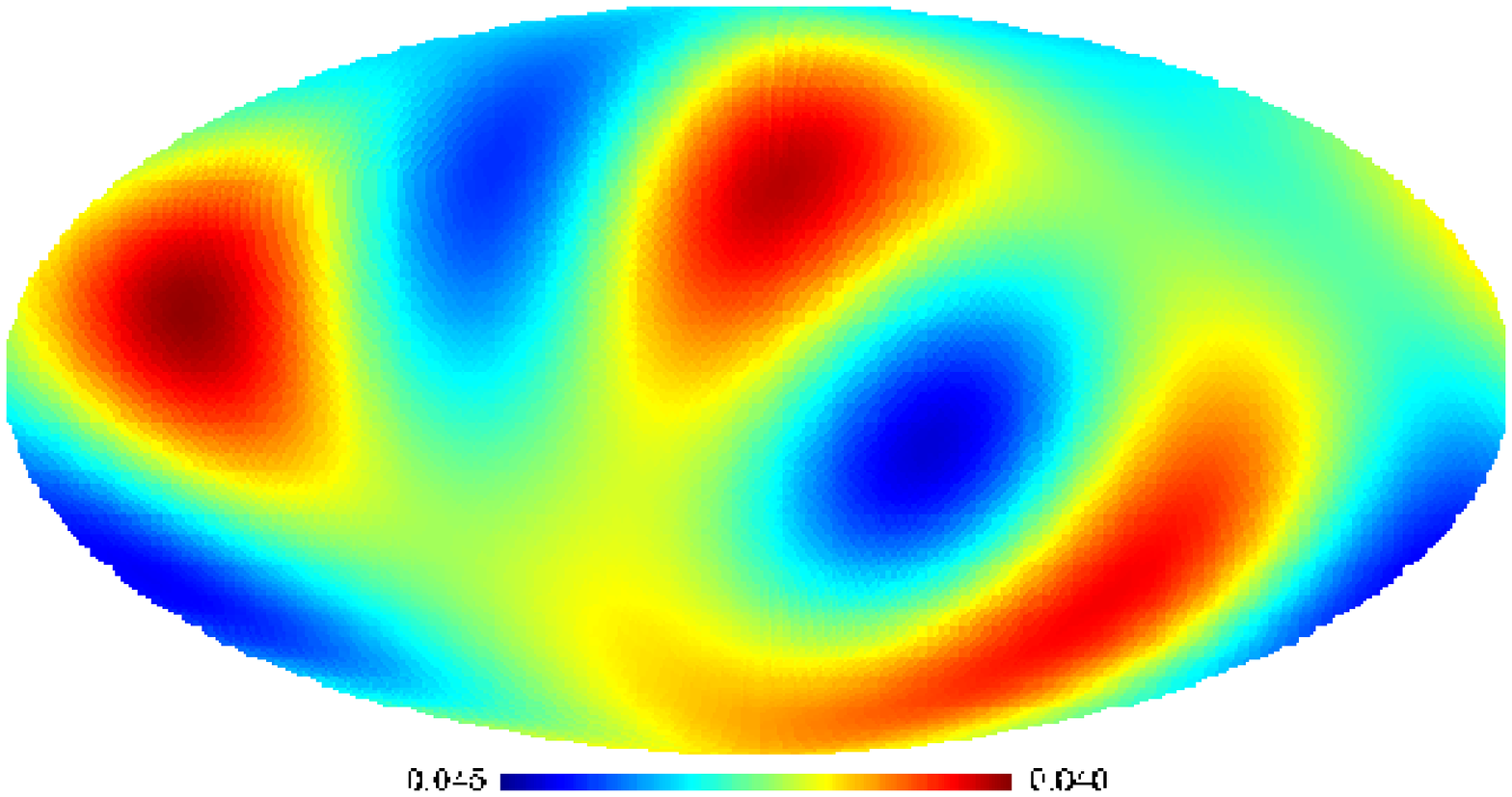,angle=90,width=5.5cm}\label{out3B}\hfill
\end{center}
\end{minipage}%
\hfill
\begin{minipage}{55mm}
\begin{center}
\psfig{file=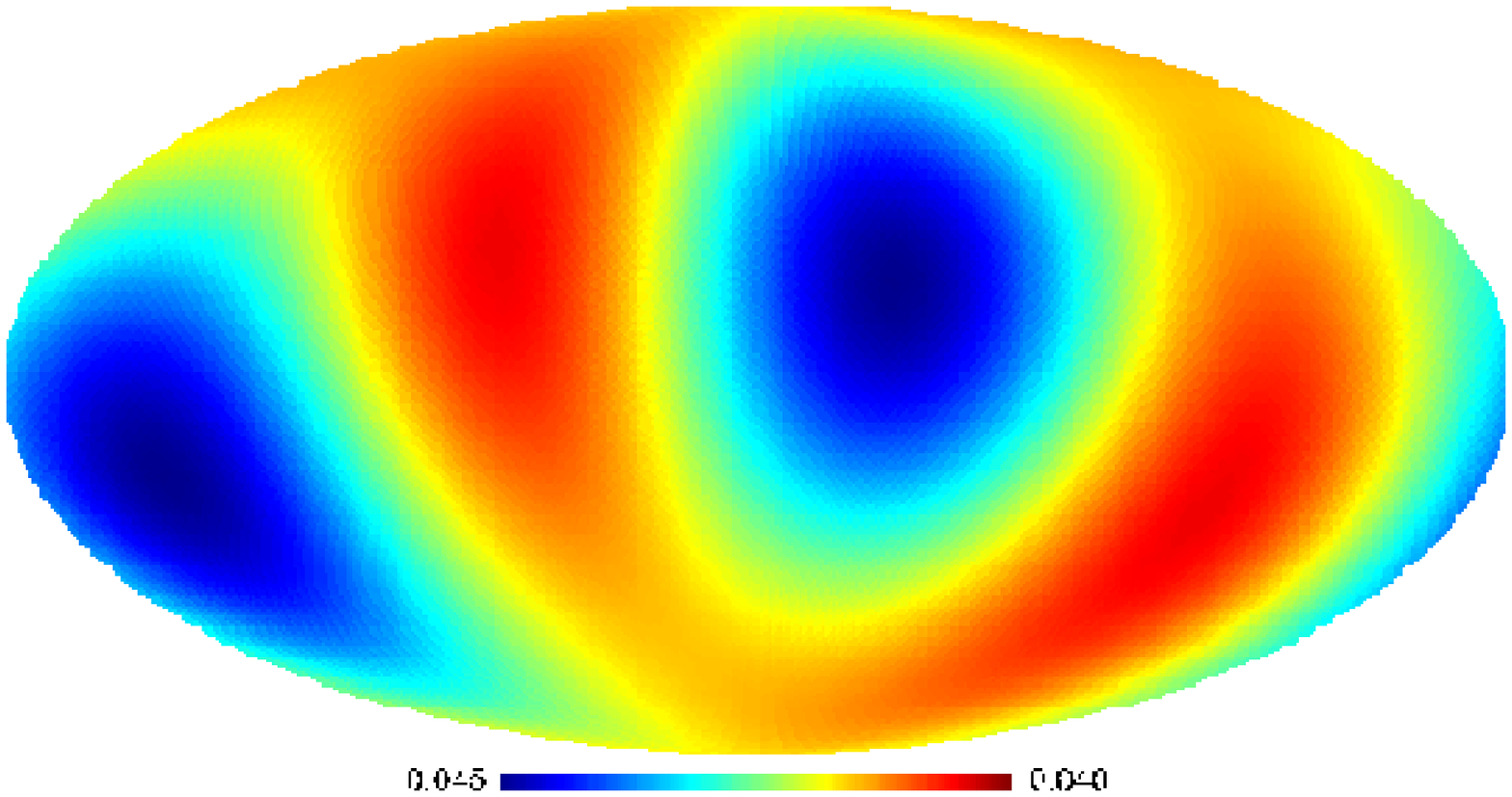,angle=90,width=5.5cm}\label{out2W}\hfill
\psfig{file=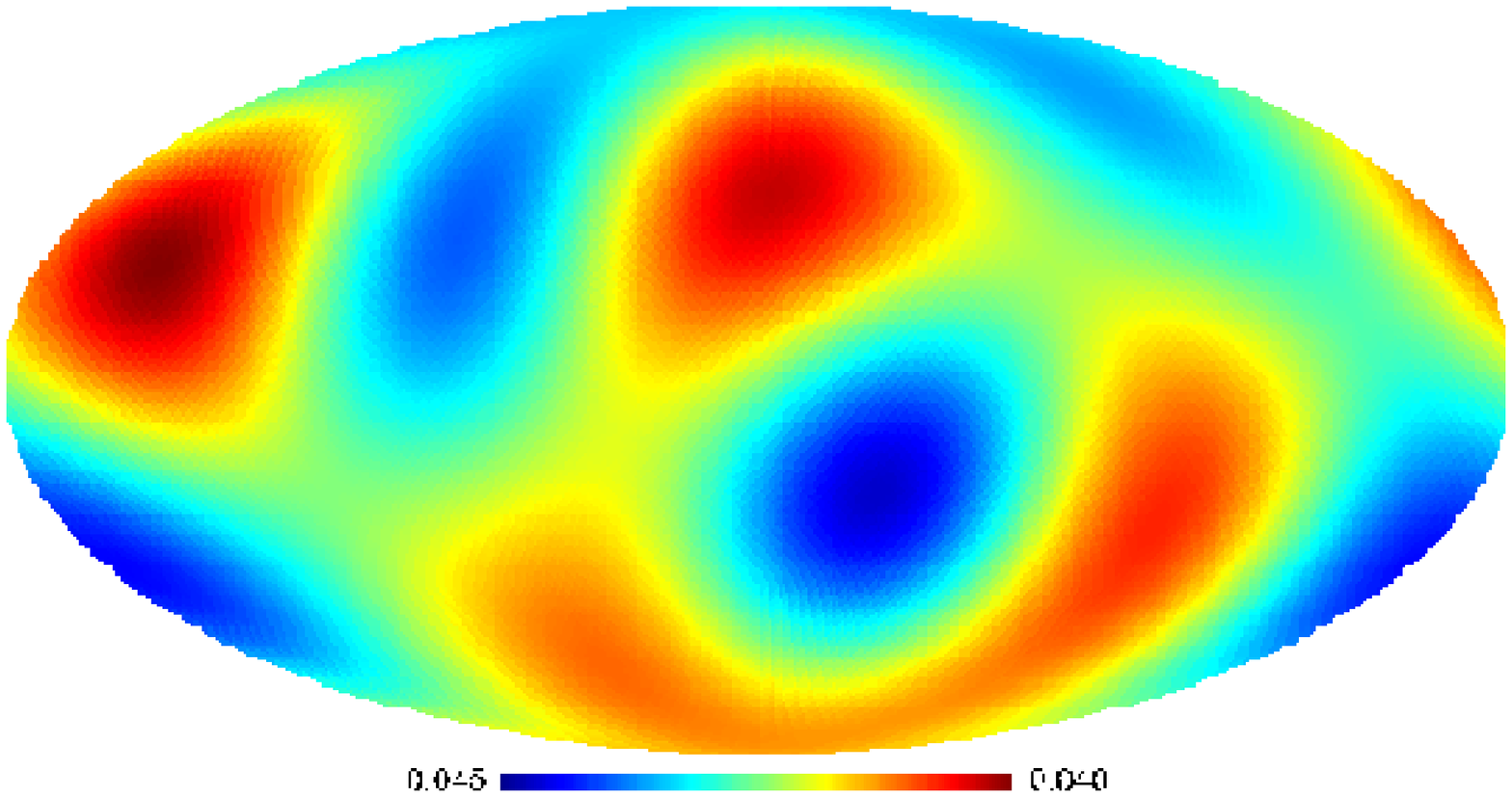,angle=90,width=5.5cm}\label{out3W}\hfill
\end{center}
\end{minipage}
\caption{The Quadrupole, $\ell=2$ (above), and the Octopole ,
$\ell=3$ (below), for the uncorrected (left) and Bianchi corrected
(middle) and the Wave corrected (right) maps. The apparent
alignment is not present in the corrected maps (mK units).}
\label{octquad}
\end{figure*}

\begin{figure*}
\centering
\begin{minipage}{75mm}
\begin{center}
\hfill \epsfig{file=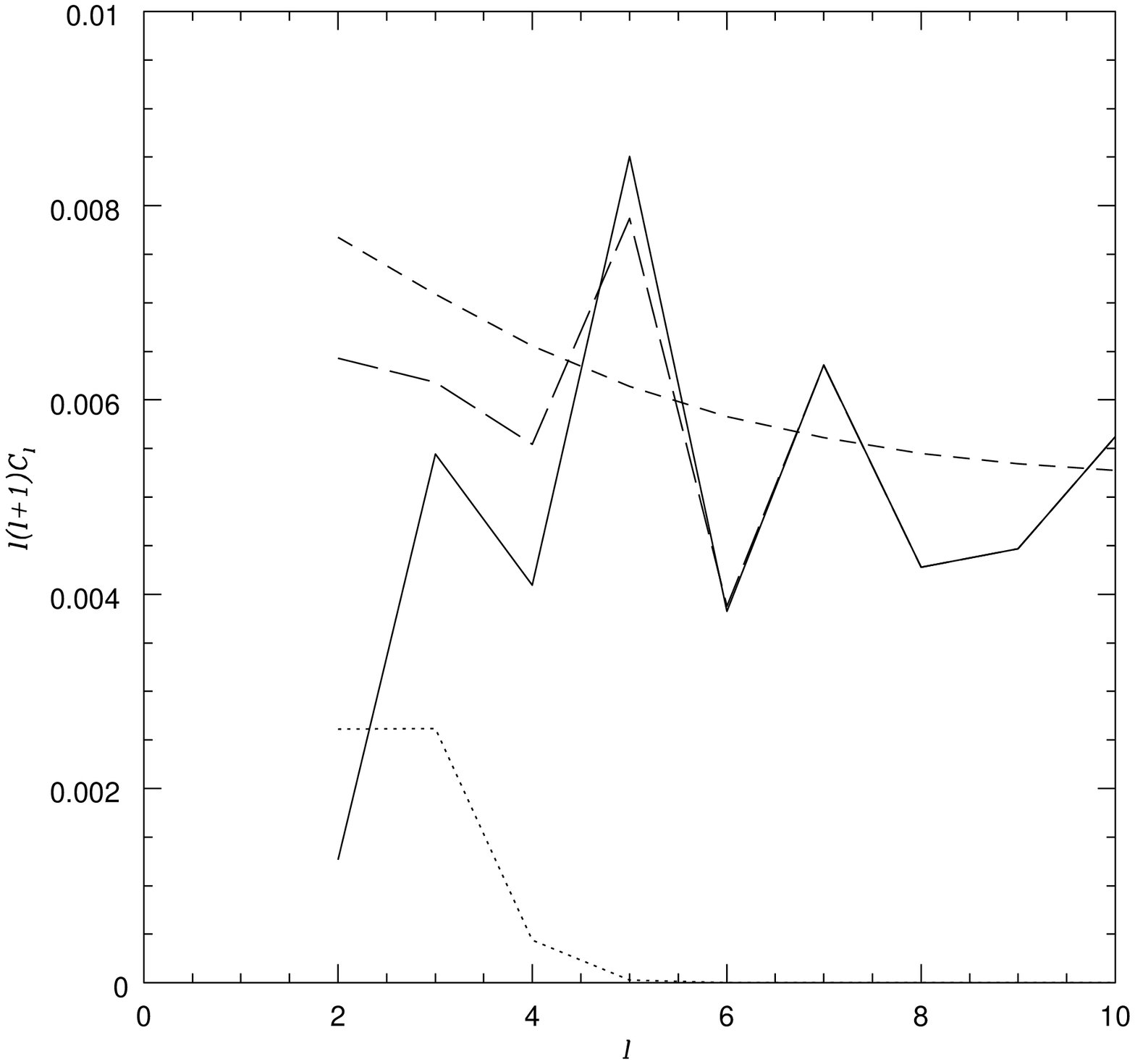,width=7.5cm}
\center{Bianchi}\hfill
\end{center}
\end{minipage}%
\hfill
\begin{minipage}{75mm}
\begin{center}
\hfill \epsfig{file=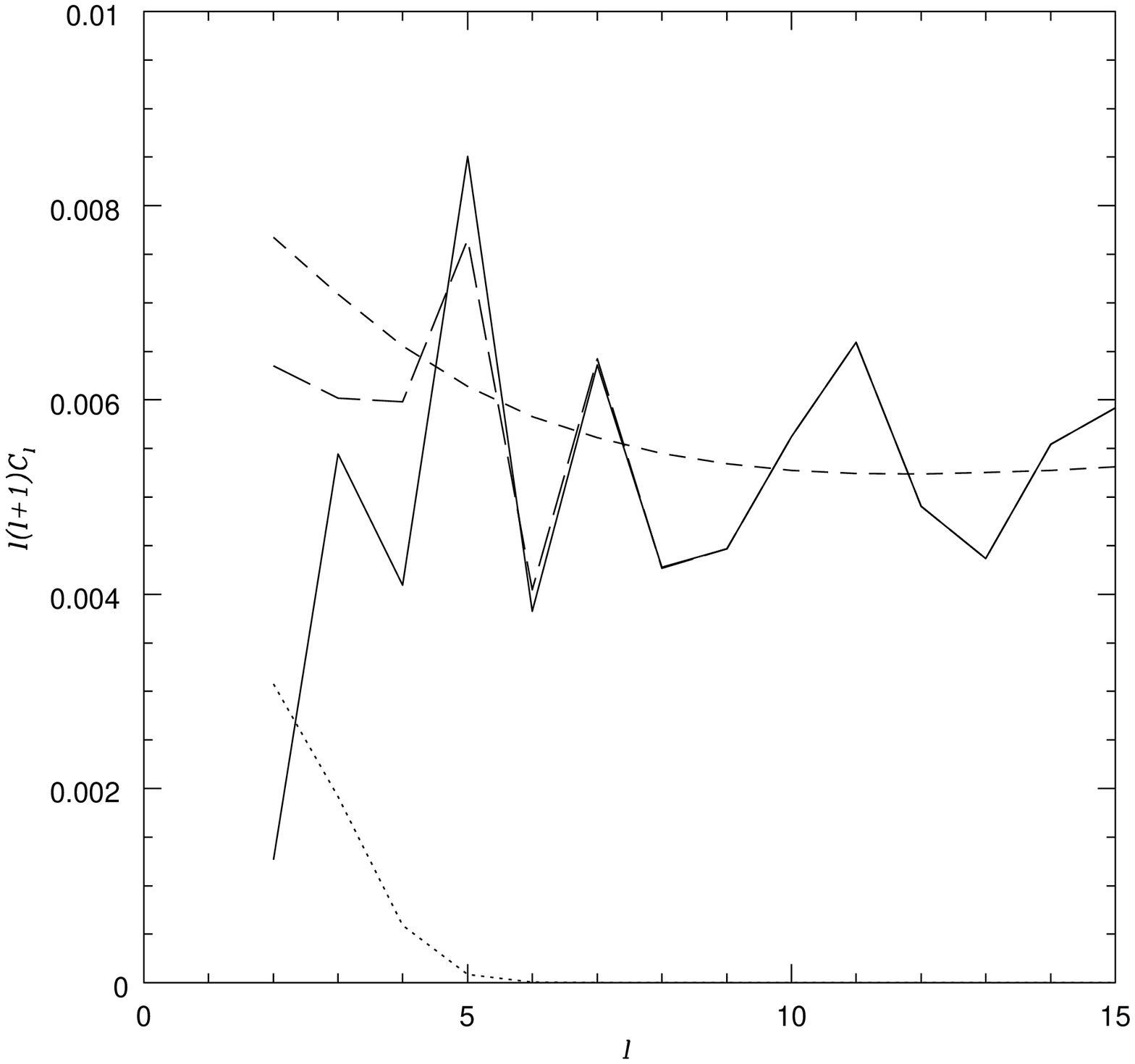,width=7.5cm} \center{Wave}\hfill
\end{center}
\end{minipage}%
\caption{The power spectrum (mK$^2$) of the map before (solid
line) and after (long-dashes) being corrected for the preferred
template (dotted). The $\Lambda$CDM theoretical power spectrum
from WMAP is also plotted (short-dashes).}\label{power}
\end{figure*}

In Figures~\ref{figB} and ~\ref{figW} we plot the map before and
after the correction, only showing the multipoles $\ell=2-5$. In
Figure~\ref{octquad} we plot the much examined Quadrupole and
Octopole before and after the corrections. The alignment
\citep{biel, teg, copilet} can clearly be seen in the before maps,
and not in the after maps, also the anomalously low power is
increased, see Figure~\ref{power}.

We have therefore proved that all known large-angle anomalies may
be coupled. There are models providing templates which
can correct, in one go, all known anomalies -- the low power and the ``axis
of evil'' effect. Whether or not the evidence for these models
is high is another matter, which we now proceed to examine.

\hfill

\section{Significance}\label{sims}
There are many ways to quantify the significance of the final
correction. As already noted in Section~\ref{intro}, the expected power
spectrum of a map will be higher than that for a purely Gaussian map if we
assume the presence of a template. Thus, when considering the
low-$\ell$ low power, if we take into account the presence of a
template, then we actually make the situation worse! We consider
simulations of Gaussian maps with the added template, for both the
Bianchi and the Wave models. We calculate power spectrum estimator
for 5000 simulations and indeed we find that the without adding a
template, only 2.6\% of simulations find a $C_2$ lower than that
observed by WMAP. And with the Bianchi template added this reduces
to 1.2\%, thus making the fit worse.

However in apparent contradiction to this we have a map that, when
corrected for the preferred template, finds excellent consistency
with the theoretical power spectrum. So, we ask a different
question here. We allow 1000 simulations to find their best
template, for both model types. We record the $\chi^2_d$ before
the correction, and the $\alpha$, $\chi^2_t$, $\Delta$ of the
correction, and we only look at those simulations that find a low
initial $\chi^2_d$ (525 of our 1000 simulations). We can now
assess the significance of our results, given the prior assumption
of a destructive alignment between the Gaussian process and the
template.

We find the simulations return similar results to the WMAP map.
That is we looked at the range of $\alpha$ values returned, and
various functions of ($\alpha, \chi^2_t, \Delta, \chi^2_d$) and
found our WMAP result to not be particularly anomalous compared to
the simulation results. This was the case for both the Bianchi and
the Wave templates.

We therefore must conclude that the detections are not
significant. However, we find the method still of interest, and
note it can be used to investigate new models of contamination in
the future. We give one example of a possible improvement.

Qualitatively the crux of the issue lies in that we must rely on a
chance anti-alignment of phases between template and Gaussian
process. We don't see power in some $m$ modes because it was there
both in the template and in the Gaussian process and the two added
destructively. This will always sound suspicious and we suggest
that models might be set up where the underlying non-Gaussian
process influences the production of Gaussian fluctuations so that
the two processes are anti-correlated. Specifically we could have
set up a likelihood where the $\chi^2$ functions for template and
Gaussian process added (rather than their $a_{\ell m}$). However
such work is beyond the scope of this paper.

\section{Discussion}\label{discuss}

We found templates from two different models that can
simultaneously explain the observed anisotropic alignments of the
low-$\ell$ multipoles and their low power. The corrected maps show
power spectra vastly more consistent with the standard $\Lambda$CDM
theoretical power spectrum~\citep{wmap} than the uncorrected map.
The scope of our work was two-fold.

Firstly, we proposed a formalism for dealing with templates
in the presence of data which has a low chisquared. In this approach
we promote the  $\chi^2$ to the status of relevant function of
the data for comparing theory and data. This is in contrast with the
the usual $a_{\ell m}$s, and reduces the number of degrees of freedom.
Furthermore the likelihood problem is then turned on its head in that
it enforces the strong prior that a Gaussian process with a power
known a priori must be present in the corrected maps. Thus this approach
generalizes the previous one in that it reduces to it for
cases where the data chisquared is too high, but leads to significantly
different results when the observed chisquared is too low.

Notice that standard methods for
template correction will lead to a corrected map with
a lower $\chi^2$. This is because it is more probable that the
presence of a template will increase the power than remove it. We
are therefore always on the back foot when trying to use a
template to explain a low power spectrum (see \cite{slosar,slosar2}
for a similar discussion). We note that as the current
data is such a bad fit, most of our templates ($\sim$90 \%)
improved the $C_2$ fit before we even imposed the second condition
of minimising $\Delta$.

The second purpose of our paper was to use this formalism to investigate
whether the low power (i.e. low chisquared) observed in the
low multipoles and the axis of evil effect might be related to each other,
and whether both could be due to an underlying anisotropic or inhomogeneous
model of the Universe. The axis of evil effect consists of the fact that
for a given $z$-axis orientation the power in a given multipole is
not distributed ``at random''among the various $m$ modes but
concentrates on planar models ($m=\pm\ell$) for $\ell=2,3$; furthermore the
axis for which this happens is roughly the same for $\ell=2,...,5$.
A priori the two effects should be coupled because the $C_\ell$ is an average
over $m$ of the variance $|a_{\ell m}|^2$. It looks as if some $m$-modes
missed their power and that if this was reinstated the low $C_\ell$
and the ``axis of evil'' anomalies would disappear. This might happen
after correcting the nefarious influence of a deterministic template
and it suggests that the template should have a strong non-planar component
along the preferred direction.

We considered Bianchi models and also inhomogeneous models where a
very strong long-wavelength wave is part of the background model.
We find the preferred parameters for these models and correct the maps.
The corrected maps are free of anomalies, in
particular their quadrupole and octupole are not planar and their
intensities not low. Therefore we stress that although the ``template''
detection are not found to be statistically significant they do correct statistically
significant anomalies.

While this paper was being finished two preprints appeared proposing
explanations for the axis of evil effect~\cite{pancake,sib}.
One of these is rather similar in spirit to ours~\cite{sib}:
the idea that a long wavelength mode is imprinted in the sky.
Even though our statistical treatment is rather different, both
papers highlight the same difficulty with such explanations: that in general
waves would add, rather than subtract power. Our suggested explanation
(that an anti-correlation might exist between the Gaussian and the deterministic
process) finds its counterpart in the model considered in~\cite{sib}
in the concept of multiplicative non-linear response. We find this
idea very interesting, but as with our idea of Gaussian/template
anti-correlation we stress that it may not be necessary. The fit
between the data and the pure Gaussian model is so bad that a fluke
anti-alignment of phases is already a good enough explanation
(even though clearly the Bayesian evidence for more complex models
will always be superior).

We are also examining the possible correlations between the CMB
anomalies and the large-scale structure and defer to
a future publication~\cite{ofer} comments on~\cite{pancake}.

{\bf Acknowledgements} We thank Anthony Banday, Kris Gorski,
Andrew Jaffe, Tesse Jaffe, Jo\~ao Medeiros, Carlo Contaldi, and
Max Tegmark for helpful comments. Our calculations made use of the
HEALPix package~\cite{healp} and were performed on COSMOS, the UK
cosmology supercomputer facility. KL is funded by PPARC.

\label{lastpage}

\end{document}